\documentclass[fleqn,12pt]{article}
\usepackage{espcrc1m}
\usepackage{graphicx}

\title{Hard-loop dynamics of non-abelian plasma instabilities}

\author{Michael Strickland
\address[HELSINKI]{Department of Physical Sciences and Helsinki Institute of Physics,\\
P.O. Box 64, FIN-00014 University of Helsinki, Finland}
}

\begin{document}

\maketitle

\abstract{

I discuss recent advances in the understanding of non-equilibrium gauge field 
dynamics in plasmas which have particle distributions which are locally 
anisotropic in momentum space.  In contrast to locally isotropic plasmas such 
anisotropic plasmas have a spectrum of soft unstable modes which are 
characterized by exponential growth of transverse (chromo)-magnetic fields at 
short times.  The long-time behavior of such instabilities depends on whether or 
not the gauge group is abelian or non-abelian. Here I will report on recent 
numerical simulations which attempt to determine the long-time behavior of an 
anisotropic non-abelian plasma within hard-loop effective theory.

}

\section{Introduction}

One of the mysteries emerging from the RHIC ultrarelativistic heavy-ion 
collision experiments is that the matter produced in the collisions seems to be 
well-described by hydrodynamic models.  In order to apply hydrodynamical models 
the chief requirement is that the stress-energy tensor be isotropic in momentum 
space.  Additionally, current hydrodynamic codes also assume that they can use 
an equilibrium equation of state to describe the time evolution of the produced 
matter.  Therefore, the success of these models suggests that the bulk matter 
produced is {\em isotropic} and {\em thermal} at very early times, $t < 1$ fm/c. 
Estimates of the isotropization and thermalization times from perturbation 
theory \cite{BottomUp}, however, indicate that the time scale for thermalization 
is more on the order of $t \sim 2-3$ fm/c.  This contradiction has led some to 
conclude that perturbation theory should be abandoned and replaced by some other 
(as of yet unspecified) calculational framework. However, it has been proven 
recently that previous perturbative estimates of the isotropization and 
equilibration times had overlooked an important aspect of nonequilibrium gauge 
field dynamics, namely the possibility of {\em plasma instabilities}.

One of the chief obstacles to thermalization in ultrarelativistic heavy-ion 
collisions is the intrinsic expansion of the matter produced.  If the matter 
expands too quickly then there will not be time enough for its constituents to 
interact before flying apart into non-interacting particles and therefore the 
system will not reach thermal equilibrium.  In a heavy-ion collision the 
expansion which is most relevant is the longitudinal expansion of the matter 
since at early times it's much larger than the radial expansion.  In the absence of 
interactions the longitudinal expansion causes the system to quickly become much 
colder in the longitudinal direction than in the transverse (radial) direction, 
$<\!p_L\!> \ll <\!p_T\!>$.  We can then ask how long it would take for 
interactions to restore isotropy in the $p_T$-$p_L$ plane.  In the bottom-up 
scenario \cite{BottomUp} isotropy is obtained by hard collisions between the 
high-momentum modes which interact via an isotropically screened gauge 
interaction.  The bottom-up scenario assumed that the underlying soft gauge 
modes responsible for the screening were the same in an anisotropic plasma as in 
an isotropic one. In fact, this turns out to be incorrect and in anisotropic 
plasmas the most important collective mode corresponds to an instability to 
transverse magnetic field fluctuations \cite{Stanislaw}.  Recent works have 
shown that the presence of these instabilities is generic for distributions 
which possess a momentum-space anisotropy \cite{CmodesPR,Arnold:2003rq} and have 
obtained the full hard-loop action in the presence of an anisotropy 
\cite{Mrowczynski:2004kv}.

Here I will discuss numerical results obtained within the last year which 
address the question of the long-time behavior of the instability evolution 
\cite{Arnold:2004ih,Rebhan:2004ur,Arnold:2005vb,Rebhan:2005re} within the hard-loop 
framework. This question is non-trivial in QCD due to the presence of non-linear 
interactions between the gauge degrees of freedom.  These non-linear 
interactions become important when the vector potential amplitudes become 
$<\!\!A\!\!>_{\rm soft} \sim p_{\rm soft}/g \sim (g p_{\rm hard})/g$, where 
$p_{\rm hard}$ is the characteristic momentum of the hard particles.  In QED 
there is no such complication and the fields grow exponentially until 
$<\!\!A\!\!>_{\rm hard} \sim p_{\rm hard}/g$ at which point the hard particles 
undergo large-angle scattering in the soft background field invalidating the 
assumptions underpinning the hard-loop effective action. Initial numerical toy 
models indicated that non-abelian theories in the presence of instabilities 
would ``abelianize'' and fields would saturate at  $<\!\!A\!\!>_{\rm hard}$ 
\cite{Arnold:2004ih}.  This picture was largely confirmed by simulations of the 
full hard-loop gauge dynamics which assumed that the soft gauge fields depended 
only on the direction parallel to the anisotropy vector and time 
\cite{Rebhan:2004ur}.  However, recent numerical studies have now included the 
transverse dependence of the gauge field and it seems that the result is then 
that the gauge field's dynamics changes its behavior from exponential to linear 
growth when its amplitude reaches the soft scale, $<\!\!A\!\!>_{\rm soft} \sim 
p_{\rm hard}$ \cite{Arnold:2005vb,Rebhan:2005re}.  This linear growth regime is 
characterized by a cascade of the energy pumped into the soft scale by the 
instability to higher momentum plasmon-like modes \cite{cascade}.  Below I will 
briefly describe the setup which is used by these numerical simulations and then 
discuss questions which remain in the study of non-abelian plasma instabilities.

\section{Discretized Hard-Loop Dynamics}%

At weak gauge coupling $g$, there is a separation of scales in hard momenta 
$|\mathbf p|=p^0$ of (ultrarelativistic) plasma constituents, and soft momenta 
$\sim g|\mathbf p|$ pertaining to collective dynamics. The effective field 
theory for the soft modes that is generated by integrating out the hard plasma 
modes at one-loop order and in the approximation that the amplitudes of the soft 
gauge fields obey $A_\mu \ll |\mathbf p|/g$ is that of gauge-covariant 
collisionless Boltzmann-Vlasov equations \cite{HTLreviews}. In equilibrium, the 
corresponding (nonlocal) effective action is the so-called hard-thermal-loop 
effective action which has a simple generalization to plasmas with anisotropic 
momentum distributions \cite{Mrowczynski:2004kv}. The resulting equations of 
motion are
\begin{eqnarray}
D_\nu(A) F^{\nu\mu} &=& -g^2 \int {d^3p\over(2\pi)^3} {1\over2|\mathbf p|} \,p^\mu\, 
						 {\partial f(\mathbf p) \over \partial p^\beta} W^\beta(x;\mathbf v) \, , \nonumber \\
F_{\mu\nu}(A) v^\nu &=& \left[ v \cdot D(A) \right] W_\mu(x;\mathbf v) \, , 
\label{eom}
\end{eqnarray}
where $f$ is a weighted sum of the quark and gluon distribution 
functions \cite{Mrowczynski:2004kv} and $v^\mu\equiv p^\mu/|\mathbf p|=(1,\mathbf v)$.

At the expense of introducing a continuous set of auxiliary fields 
$W_\beta(x;\mathbf v)$ the effective field equations are local.  These equations 
of motion are then discretized in space-time and ${\mathbf v}$, and solved 
numerically.  The discretization in ${\mathbf v}$-space corresponds to including 
only a finite set of the auxiliary fields $W_\beta(x;\mathbf v_i)$ with $1 \leq 
i \leq N_W$. For details on the precise discretizations used see 
Refs.~\cite{Arnold:2005vb,Rebhan:2005re}.

\begin{figure}[t]
\vspace{4mm}
\centerline{
\includegraphics[width=9.1cm]
{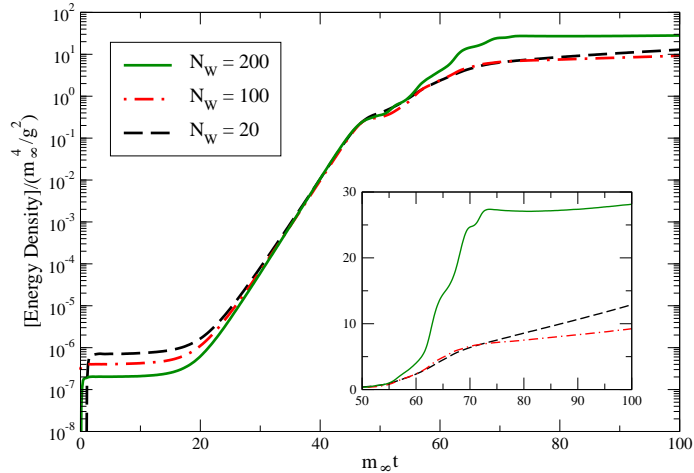}
}
\vspace{-7mm}
\caption{
Comparison of the energy transferred from hard to soft scales, $\mid\!\!\mathcal 
E({\rm HL})\!\!\mid$, for 3+1-dimensional simulations with $N_w=20,100,200$ 
on $96^3, 88^3, 69^3$ lattices. Inset shows late-time behavior on a linear 
scale.
\label{3dHLfig}}
\end{figure}

\section{Results and Discussion}

During the process of instability growth the soft gauge fields get the energy 
for their growth from the hard particles.  In an abelian plasma this energy 
grows exponentially until the energy in the soft field is of the same order of 
magnitude as the energy remaining in the hard particles.  As mentioned above in 
a non-abelian plasma one must rely on numerical simulations due to the presence 
of strong gauge field self-interactions. In Fig.~(\ref{3dHLfig}) I have plotted 
the time dependence of the energy extracted from the hard particles obtained in 
a 3+1 dimensional simulation of an anisotropic plasma initialized with very weak 
random color noise \cite{Rebhan:2005re}.  As can be seen from this figure at 
$m_\infty t \sim 60$ there is a change from exponential to linear growth with the 
late-time linear slope decreasing as $N_W$ is increased.

The first conclusion that can be drawn from this result is that within non-abelian 
plasmas instabilities will be less efficient at isotropizing the plasma 
than in abelian plasmas.  However, from a theoretical perspective ``saturation'' 
at the soft scale implies that one can still apply the hard-loop effective 
theory self-consistently to understand the behavior of the system at late times. 
Looking forward, I note that the latest simulations 
\cite{Arnold:2005vb,Rebhan:2005re} have only presented results for distributions 
with a finite ${\mathcal O}(1-10)$ anisotropy and these seem to imply that in 
this case the induced instabilities will not have a significant effect on the hard 
particles.  This means, however, that due to the continued expansion of the 
system that the anisotropy will increase.  It is therefore important to 
understand the behavior of the system for more extreme anisotropies. 
Additionally, it would be very interesting to study the hard-loop dynamics in an 
expanding system.  Naively, one expects this to change the growth from 
$\exp(\tau)$ to $\exp(\sqrt\tau)$ at short times but there is no clear 
expectation of what will happen in the linear regime.  The short-time picture 
has been confirmed by early simulations of instability development in an 
expanding system of classical fields \cite{paulnew}.  It would therefore be 
interesting to incorporate expansion in collisionless Boltzmann-Vlasov transport 
in the hard-loop regime and study the late-time behavior in this case.

I note in closing that the application of this framework to phenomenologically 
interesting couplings is suspect since the results obtained strictly only apply 
at very weak couplings; however, the success of hard-thermal-loop perturbation 
theory at couplings as large as $g \sim 2$ \cite{HTLpt} suggests that the 
nonequilibrium hard-loop theory might also apply at these large couplings.  For 
going to even larger couplings perhaps colored particle-in-cell simulations 
\cite{Dumitru:2005gp} could be used if they are extended to full 3+1 dynamics.

\section*{Acknowledgements}
I would like to thank my collaborators A.~Rebhan and P.~Romatschke.
This work was supported by the Academy of Finland, contract no. 77744.

\end{document}